\documentclass[10pt]{article}%

\usepackage{amsmath}
\usepackage{amsfonts}
\usepackage[psamsfonts]{amssymb}
\usepackage{array}
\usepackage{graphicx}
\usepackage{lscape}
\usepackage[dvips]{color}

\newcommand{\h}{h}
\newcommand{\longquad}{\qquad\qquad\qquad\qquad\qquad\qquad\qquad\qquad}

\title{Reparametrization invariance of the classical metric}

\author{G.G. Kirilin\footnote{G.G.Kirilin@inp.nsk.su}\\
\textit{Budker Institute of Nuclear Physics SB RAS, Novosibirsk,
Russia}}

\begin{document}

\maketitle

\begin{abstract}
There is a statement on the parametrization dependence of the
classical metric in the recent paper of N.E.J. Bjerrum-Bohr, J.F.
Donoghue, B.R. Holstein, gr-qc/0610096. I completely disagree with
this statement. Here I show reparametrization invariance of the
classical metric.
\end{abstract}

\noindent\rule{\textwidth}{.1mm}

\section{General consideration of the reparametrization transformation}

The statement made in the paper \cite{BDH1} is that using the
following parametrization of the metric
\begin{align}
\overline{g}_{\mu\nu}=\eta_{\mu\nu}+\h_{\mu\nu}-\frac{a}{4}\,\h_{\mu\alpha}\h_{\nu
}^{\alpha}\,,\label{e1}\,%
\end{align}
to solve the Einstein equations, where $a$ is an arbitrary
constant, one derives the metric, which depends on $a$.

Let us consider this statement in detail. Expansion of the
Schwarzschild metric in $r_g/r$ can be represented in the form
\begin{align}
\overline{g}_{\mu\nu}=\eta_{\mu\nu}+\underline{g}{}_{\mu\nu}
+\underline{\underline{g}}{}_{\mu\nu} +
\mathcal{O}(r^3_g/r^3)\label{e3},
\end{align}
where $\underline{g}{}_{\mu\nu}$ and
$\underline{\underline{g}}{}_{\mu\nu}$ are the first and the
second order corrections, respectively. In harmonic gauge,
\begin{align}
\partial_\mu(\sqrt{-\overline{g}}\,\overline{g}{}^{\mu\nu})=0,\label{e3a}
\end{align}
the corrections of first order have the form
\begin{align}
\underline{g}{}_{00}=-\frac{r_{g}}{r},\quad\underline{g}{}_{ij}=-\frac{r_{g}}{r}\,%
\delta_{ij}\,.\label{e4}%
\end{align}
Using these corrections to find a perturbative solution of the
Einstein equations, one can derive the second order corrections:
\begin{align}
\underline{\underline{g}}{}_{00}=\frac{r_{g}^{2}}{2r^{2}},\quad\underline
{\underline{g}}{}_{ij}=-\frac{r_{g}^{2}}{4r^{2}}\left(  \delta^{ij}+n^{i}%
n^{j}\right).\label{e5}
\end{align}
It follows from the expression (\ref{e1}) that
$\underline{g}{}_{\mu\nu}=\underline{\h}{}_{\mu\nu}$, where
$\underline{\h}{}_{\mu\nu}$ is the correction of first order in
$r_g/r$ to $\h_{\mu\nu}$.

The first aspect to be clarified is that the parametrization of
metric perturbations does not change the gauge conditions for
total metric. If one uses the harmonic gauge
$\partial_\mu(\sqrt{-\overline{g}}\,\overline{g}{}^{\mu\nu}(h_{\alpha\beta}))$=0,
where
$\sqrt{-\overline{g}}\,\overline{g}{}^{\mu\nu}(h_{\alpha\beta})$
is an analytical function of $h_{\alpha\beta}$, then after
reparametrization $h_{\alpha\beta}\to f(h_{\alpha\beta})$, where
$f(h_{\alpha\beta})$ is an analytical function of
$h_{\alpha\beta}$, the gauge conditions become
$\sqrt{-\overline{g}}\,\overline{g}{}^{\mu\nu}(f(h_{\alpha\beta}))=0$.
The gauge conditions for \textit{total} metric (\ref{e3a}) are the
same, whatever parametrization for perturbations is used.

The statement of the paper \cite{BDH1} can be reduced to the
following: if, in the Einstein equations, one uses the following
expansion
\begin{align}
\overline{g}{}_{\mu\nu}=\eta_{\mu\nu}+\underline{g}{}_{\mu\nu}+\left(
\underline{\underline{h}}{}_{\mu\nu}+w_{\mu\nu}\right)+
\mathcal{O}(r^3_g/r^3),\label{e6}
\end{align}
instead of Eq.\,\ref{e3}, where $w_{\mu\nu}$ is the known function
\begin{align}
w_{\mu\nu}=-\frac{a}{4}\,\underline{h}{}_{\mu\alpha}\underline{h}{}_{\nu}^{\alpha}
=-\frac{a}{4}\,\underline{g}{}_{\mu\alpha}\underline{g}{}_{\nu}^{\alpha}\,,\label{e7}
\end{align}
finds $\underline{\underline{h}}{}_{\mu\nu}$ and inserts it into
Eq.\,\ref{e6}, then one derives the corrections of second order
which differ from those of Eq.\,\ref{e5}.

The statement can be repeated once more in a different form.
Assume that one can find $\underline{\underline{g}}{}_{\mu\nu}$
from the equation
\begin{align}
F(\underline{\underline{g}}{}_{\mu\nu})=0\,,\label{e8}
\end{align}
where $F$ is a certain functional. Let us substitute
$\underline{\underline{g}}{}_{\mu\nu}$ by
$\underline{\underline{\h}}{}_{\mu\nu}+w_{\mu\nu}$ in this
equation, where $w_{\mu\nu}$ is a known function and
$\underline{\underline{\h}}{}_{\mu\nu}$ is a new function to be
found:
\begin{align}
F(\underline{\underline{\h}}{}_{\mu\nu}+w_{\mu\nu})=0.\label{e9}
\end{align}
The statement of the paper \cite{BDH1} is equivalent to the
following one: if one finds $\underline{\underline{\h}}$ from the
equation (\ref{e9}), then
$\underline{\underline{\h}}{}_{\mu\nu}+w_{\mu\nu}$ would be
different from the solution of the equation (\ref{e8}). It is
obviously incorrect. In the following section, I will demonstrate
the complete series of the arithmetical mistakes that led the
authors of Ref.\,\cite{BDH1} to the false conclusions.

\section{Solution of the Einstein equations}

The first incorrect formula in the paper \cite{BDH1} is (Eq.\,4 of
Ref.\,\cite{BDH1})
\begin{align}
h_{\mu\nu}(x)=-16\pi G\int d^{3}y\,D(x-y)\left(  T_{\mu\nu}\left(
y\right) -\frac{1}{2}\eta_{\mu\nu}T\left(  y\right)
\right).\label{e9a}
\end{align}
I would like to stress here that the authors of Ref.\cite{BDH1}
use harmonic gauge (\ref{e3a}) and they derive the second order
corrections from Eq.\ref{e9a}. Putting $a=0$ yields
$\underline{\underline{\h}}{}_{\mu\nu} =
\underline{\underline{g}}{}_{\mu\nu}$, but, as follows from
Eq.\,\ref{e9a},
\begin{align}
\partial_{\mu}\left(  \underline{\underline{h}}^{\mu\nu}-\frac{1}{2}\eta
^{\mu\nu}\underline{\underline{h}}\right)  =\partial_{\mu}\left(
\underline{\underline{g}}^{\mu\nu}-\frac{1}{2}\eta^{\mu\nu}\underline
{\underline{g}}\right)  =0,\label{e10}
\end{align}
which is obviously incorrect (see the explicit form of
$\underline{\underline{g}}{}_{\mu\nu}$ in Eq.\,\ref{e5}),
consequently the expression (\ref{e9a}) is incorrect. The correct
formula is Eq.\,(A12) given in the paper \cite{BDH2} by the same
authors. It is useful to write this equation in the form
\begin{align}
& \square\left(
\underline{\underline{g}}{}_{\mu\nu}-\frac{1}{2}\eta_{\mu\nu
}\underline{\underline{g}}\right)
-2P_{\mu\nu}^{\lambda\sigma}\partial _{\lambda}\left(
\partial^{\beta}\underline{\underline{g}}_{\beta\sigma
}-\frac{1}{2}\partial_{\nu}\underline{\underline{g}}\right)
=-16\pi G T_{\mu\nu}^{grav}(\underline{g})\label{e11}\,,\\%
& P_{\mu\nu}^{\lambda\sigma}=\frac{1}{2}\left(  \delta_{\mu}^{\lambda}%
\delta_{\nu}^{\sigma}+\delta_{\nu}^{\lambda}\delta_{\mu}^{\sigma}%
-\eta^{\lambda\sigma}\eta_{\mu\nu}\right)\label{e12},
\end{align}
where $T_{\mu\nu}^{grav}(\underline{g})$ is defined by the
expression (A14) of Ref\,\cite{BDH2}. I have to write out the
expression (\ref{e11}) because the corresponding formula (A13) in
Ref.\,\cite{BDH2} is incorrect. However, it was pointed out
correctly that (in the case when $a=0$) the expression (\ref{e11})
can be reduced to formula (A16) of Ref.\,\cite{BDH2}.

It is easy to derive the analogue of the formula (A16) of
Ref.\,\cite{BDH2} in the case when $a\neq 0$. Actually, the
substitution
$\underline{\underline{g}}{}_{\mu\nu}=\underline{\underline{\h}}{}_{\mu\nu}+w_{\mu\nu}$
in the equation (\ref{e11}) yields
\begin{align}
& \square\left(
\underline{\underline{\h}}{}_{\mu\nu}-\frac{1}{2}\eta_{\mu\nu
}\underline{\underline{\h}}\right)
-2P_{\mu\nu}^{\lambda\sigma}\partial _{\lambda}\left(
\partial^{\beta}\underline{\underline{\h}}{}_{\beta\sigma
}-\frac{1}{2}\partial_{\sigma}\underline{\underline{\h}}\right)
=-16\pi G
\left(T_{\mu\nu}^{grav}(\underline{g})+\tilde{T}_{\mu\nu}(w)\right)\label{e12a}\,,%
\end{align}
where
\begin{align}
\tilde{T}_{\mu\nu}=\frac{1}{16\pi G}\left(  \square\left(
w_{\mu\nu}-\frac {1}{2}\eta_{\mu\nu}w\right)
-2P_{\mu\nu}^{\lambda\sigma}\partial_{\lambda }\left(
\partial^{\beta}w_{\beta\sigma}-\frac{1}{2}\partial_{\sigma}w\right)
\right)\,.\label{e13}
\end{align}

It is easy to notice that $\tilde{T}_{\mu\nu}$ is exactly the
$a$-dependent correction calculated in \cite{BDH1}, formula~(3) of
Ref.\,\cite{BDH1}. In fact, $\partial_\mu\tilde{T}^{\mu}_{\nu} =0$
for \textit{any} $w_{\mu\nu}$ by virtue of the contracted Bianchi
identity\footnote{the following identity
$\eta^{\alpha\nu}\partial_{\alpha}(\underline{R}{}_{\mu\nu}-\eta_{\mu\nu}\underline{R}/2)=0$
is meant, where $\underline{R}{}_{\mu\nu}$ is the first order term
in the expansion of the Ricci tensor in $h_{\mu\nu}$}, therefore,
this correction can be included in the transverse form factors
(Eq.\,2 of Ref.\,\cite{BDH1}). $\tilde{T}_{\mu\nu}$ can contribute
only to the form factor $F_2(q^2)$ in the formula (2) of
Ref.\,\cite{BDH1} because $w_{\mu\nu}$ consists of the product of
the fields $h_{\mu\nu}$ in our particular case (\ref{e7}). It is
easy to see that $\tilde{T}_{\mu\nu}$ is the correction induced by
the structure $a \mathcal{X}^{\alpha\beta}_{\gamma\delta}$ in my
comment \cite{kk3}.

To simplify the term, that is proportional to
$P_{\mu\nu}^{\lambda\sigma}$ in the lhs of the equation
(\ref{e12a}), we insert the definition (\ref{e1}) in the gauge
condition
$\partial_\mu(\sqrt{-\overline{g}}\,\overline{g}{}^{\mu\nu})=0$.
As a result, we have:
\begin{align}
0  & =-\left(
\partial_{\mu}\underline{h}^{\mu\nu}-\frac{1}{2}\,\partial
^{\nu}\underline{h}\right)  \\
& =-\left(  \partial_{\mu}\underline{\underline{h}}^{\mu\nu}-\frac{1}%
{2}\,\partial^{\nu}\underline{\underline{h}}\right)
+\frac{a}{4}\partial
_{\mu}P_{\gamma\delta}^{\mu\nu}\underline{h}_{\alpha}^{\gamma}\underline
{h}^{\alpha\delta}+\underline{h}^{\mu\sigma}\left(  \partial_{\mu}%
\underline{h}_{\sigma}^{\nu}-\frac{1}{2}\,\partial^{\nu}\underline{h}_{\mu
\sigma}\right)\label{e14}.
\end{align}
Using Eq.\,\ref{e14} yields
\begin{align}
\square\left(
\underline{\underline{h}}{}_{\mu\nu}-\frac{1}{2}\eta_{\mu\nu
}\underline{\underline{h}}\right)    & =-16\pi
GT_{\mu\nu}+2P_{\mu\nu }^{\lambda\sigma}\partial_{\lambda}\left(
\underline{h}^{\alpha\beta}\left(
\partial_{\alpha}\underline{h}{}_{\beta\sigma}-\frac{1}{2}\partial_{\sigma
}\underline{h}{}_{\alpha\beta}\right)  \right)\notag\\
&\longquad+\frac{a}{2}P_{\mu\nu}%
^{\lambda\sigma}\partial_{\lambda}\partial_{\beta}P^{\beta}{}_{\sigma
,\gamma\delta}\underline{h}{}_{\alpha}^{\gamma}\underline{h}^{\alpha\delta}\,,\label{e15}%
\end{align}
where $T_{\mu\nu}  =T_{\mu\nu}^{grav}+\tilde{T}_{\mu\nu}$.

Using the exact form of
$\underline{h}{}_{\mu\nu}=\underline{g}{}_{\mu\nu}$ (see
Eq.\,\ref{e4}), we transform Eq.\,\ref{e15} to
\begin{align}
\square\underline{\underline{h}}{}_{\mu\nu}=-16\pi G\left(
T_{\mu\nu}-\frac {1}{2}\eta_{\mu\nu}T\right) -\partial_{\mu}\left(
f\,\partial_{\nu}f\right) -\partial_{\nu}\left(
f\,\partial_{\mu}f\right)
-\frac{a}{2}\,\partial_{\mu }\partial_{\nu}\,f\,^{2}\,,\label{e15a}%
\end{align}
where $f=-r_g/r$. The expression (\ref{e15a}) is the analogue of
the formula (A16) of Ref.\cite{BDH2}, for an arbitrary $a$.

Solution of the equation (\ref{e15a}) has the form
\begin{align}
\underline{\underline{h}}{}_{00}  & =\left(  2+a\right)  \,\frac{r_{g}^{2}%
}{4r^{2}},\label{e16}\\
\underline{\underline{h}}{}_{ij}  &
=-\frac{r_{g}^{2}}{4r^{2}}\left( \delta_{ij}+n_{i}n_{j}\right)
+\left[ -\frac{r_{g}^{2}}{4r^{2}}\,a\,\left(
3\delta_{ij}-4n_{i}n_{j}\right)  \right]
+\frac{r_{g}^{2}}{4r^{2}}\,2a\left(
\delta_{ij}-2n_{i}n_{j}\right)  \label{e17}\\
& =-\frac{r_{g}^{2}}{4r^{2}}\left(  \delta_{ij}+n_{i}n_{j}\right)
-\frac{a}{4}\,\frac{r_{g}^{2}}{r^{2}}\,\delta_{ij}.\label{e18}
\end{align}
The correction (\ref{e16}) is derived in Ref.\,\cite{kk3}. In
expression (\ref{e17}) we separate by the square brackets the term
which the authors of Ref.\,\cite{BDH1} would have obtained from
the expression (\ref{e9a}) if they had made the Fourier transform
correctly.

Inserting
$h_{\mu\nu}=\underline{h}_{\mu\nu}+\underline{\underline{h}}{}_{\mu\nu}$
in the equation (\ref{e1}), we find the corrections (\ref{e4}) and
(\ref{e5}) exactly.

\section{Conclusion}

Reparametrization invariance of the classical metric is a
triviality. It is important to understand that a parametrization
is only a \textit{way} of calculation. All conclusions of the
paper \cite{BDH1} are based on the mistaken calculations,
therefore, they are incorrect as a whole.

A nontrivial requirement appears  when one considers quantum
corrections to the Schwarzschild metric \cite{kk3}. After
averaging over the short-wavelength modes of gravitational field
one derives an effective action (a non-local effective action, in
general case) which depends only on long-range modes of
gravitational field. All information about the short-wavelength
fluctuations (i.e. their parametrization or even their gauge) is
lost after these fluctuations are integrated out (after the
averaging).

Following are the several points concerning terminology. Since the
corrections to $h_{\mu\nu}$ are not the corrections to the metric
(but related to them by the "complicated"\ equation (\ref{e1})), I
would not even call the quantity $T^{\mu\nu}$ (in the rhs of
Eq.\,\ref{e15a}) "energy momentum tensor for the gravitational
field". Correspondingly, $\tilde{T}^{\mu\nu}$ (see Eq.\,\ref{e13}
) is not a correction to the energy momentum tensor, because
$w_{\mu\nu}$ can be any arbitrary function which is not related to
the source of gravitational field. Terminology is matter of taste
though. In any case one can see that $\tilde{T}^{\mu\nu}$ is a
total derivative and the statement of the paper \cite{BDH1} that
"the amount of energy and momentum that carried in the classical
field also varies with the parametrization" is incorrect.

\bibliographystyle{amsplain}

\begin{thebibliography}{99}
%
\bibitem{BDH1} N.E.J. Bjerrum-Bohr, J.F. Donoghue, B.R. Holstein, gr-qc/0610096.
%
\bibitem{BDH2} N.E.J. Bjerrum-Bohr, J.F. Donoghue, B.R. Holstein, Quantum corrections to the Schwarzschild
and Kerr metric, Phys. Rev. \textbf{D68}, (2003) 084005;
hep-th/0211071.
%
\bibitem{kk3} G.G. Kirilin, Quantum corrections to the Schwarzschild metric
and reparametrization transformations; gr-qc/0601020
%
\end{thebibliography}

\end{document}